\documentclass{pkas}

\def\year{2014} 
\def\month{March} 
\def\volume{00} 
\def\issue{0} 
\def\beginpage{1} 
\def\endpage{99} 
\def\received{---} 
\def\accepted{---} 
\date{Received \received ; accepted \accepted}

\title{
The AKARI Deep Field South:
Pushing to High Redshift
}

\author[1]{David L. Clements}

\affil[1]{Blackett Laboratory, Physics Department, Imperial College, Prince Consort Road, London SW7 2AZ, UK; \email{d.clements@imperial.ac.uk}}

\def\corrauthor{
David L. Clements
}

\def\runningauthor{
David L. Clements
}

\def\runningtitle{
The AKARI Deep Field South
}

\def\keywords{
astronomy --- astrophysics --- journals: individual: PKAS
}

\def\abstracttext{
The AKARI Deep Field South (ADF-S) is a large extragalactic survey field that is covered by multiple instruments, from optical to far-IR and radio. I summarise recent results in this and related fields prompted by the release of the {\em Herschel} far-IR/submm images, including studies of cold dust in nearby galaxies, the identification of strongly lensed distant galaxies, and the use of colour selection to find candidate very high redshift sources. I conclude that the potential for significant new results from the ADF-S is very great. The addition of new wavelength bands in the future, eg. from {\em Euclid}, SKA, ALMA and elsewhere, will boost the importance of this field still further.
}

\begin{document}
\pkashead 


\section{Introduction}

The quest for rare objects, be they extreme objects at high redshift or unusual in some other way, requires surveys of large areas of the sky using instruments that operate at a range of different wavelengths. The definition of what constitutes `large area' and a `range of wavelengths' will differ depending on the scientific goals of the survey. The large investment of time, both observing time on telescopes and researcher time in reducing and analysing the data, generally means that the production of such surveys requires significant resources. Often the data in these surveys is rapidly made publicly available so that the investment of resources can produce as great a scientific return as possible. The modern day exemplar of such a project is perhaps the original Hubble Deep Field (HDF) programme (Williams et al., 1996). In this case `large area' meant just a few square arcminutes and `range of wavelengths' meant just four optical passbands, but the scientific impact of these observations was huge.

Nearly twenty years later and our views of what constitute `large area' and `range of wavelengths' have expanded considerably. We now have surveys that cover wavelengths from X-Ray to radio in multiple bands and in fields that are tens to hundreds of square degrees in size. In the future we will have very large fractions of the sky observed to depths and resolutions comparable to those achieved in the HDF thanks to the {\em Euclid} mission (Laureijs et al., 2010) and the Large Synoptic Survey Telescope (LSST, Ivezic et al., 2008), and much more besides. Even in this future of large surveys, though, regions that are covered by a very wide range of bands will still be unusual. In this context the AKARI Deep Field South (ADF-S) will remain a very useful resource since it has a wide area (about 10 sq. deg.) and coverage from the optical to the radio in multiple bands (see eg. Clements, 2012, for a list). The field is also well placed, at RA=4:44:00 Dec=-53:20:00 to benefit from future southern instruments such as SKA and ALMA, and, since it lies close to the southern ecliptic pole, it will be well covered by space missions such as {\em Euclid} and {\em SPICA} (Nakagawa et al., 2015).

\section{ADF-S from {\em Herschel}}

\begin{figure*}
\centering
\includegraphics[width=140mm]{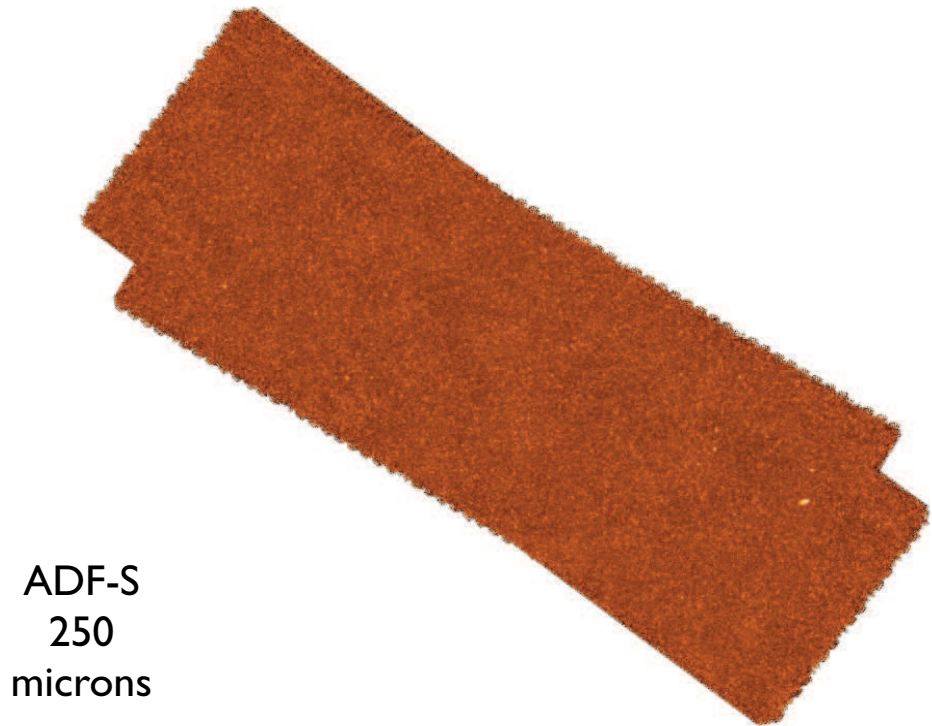}
\caption{The ADF-S as seen by {\em Herschel} at 250 $\mu$m.}
\label{fig:adfs}
\end{figure*}

The latest addition to the wide range of publicly available data on the ADF-S are the 250, 350 and 500$\mu$m maps of the region produced by the {\em Herschel Space Observatory} (Pilbratt et al., 2010) as part of the HerMES survey (Oliver et al., 2012) using the SPIRE instrument (Griffin et al., 2010). The SPIRE 250$\mu$m image is shown in Fig. \ref{fig:adfs}. When these observations are combined with the existing {\em AKARI} and {\em Spitzer} data, the sources in this entire $\sim$10 sq. deg. field are covered from the mid- to far-IR over more than a decade in wavelength, from 25 to 500 $\mu$m, albeit at varying sensitivities and angular resolutions. PACS observations at 100 and 160$\mu$m (Poglitsch et al., 2010) were also obtained by the HerMES survey which, while comparable in sensitivity to the {\em AKARI} (Shirahata et al., 2009) and {\em Spitzer}  data (Clements et al., 2011), are at significantly higher resolution.

\begin{figure*}
\centering
\includegraphics[width=140mm]{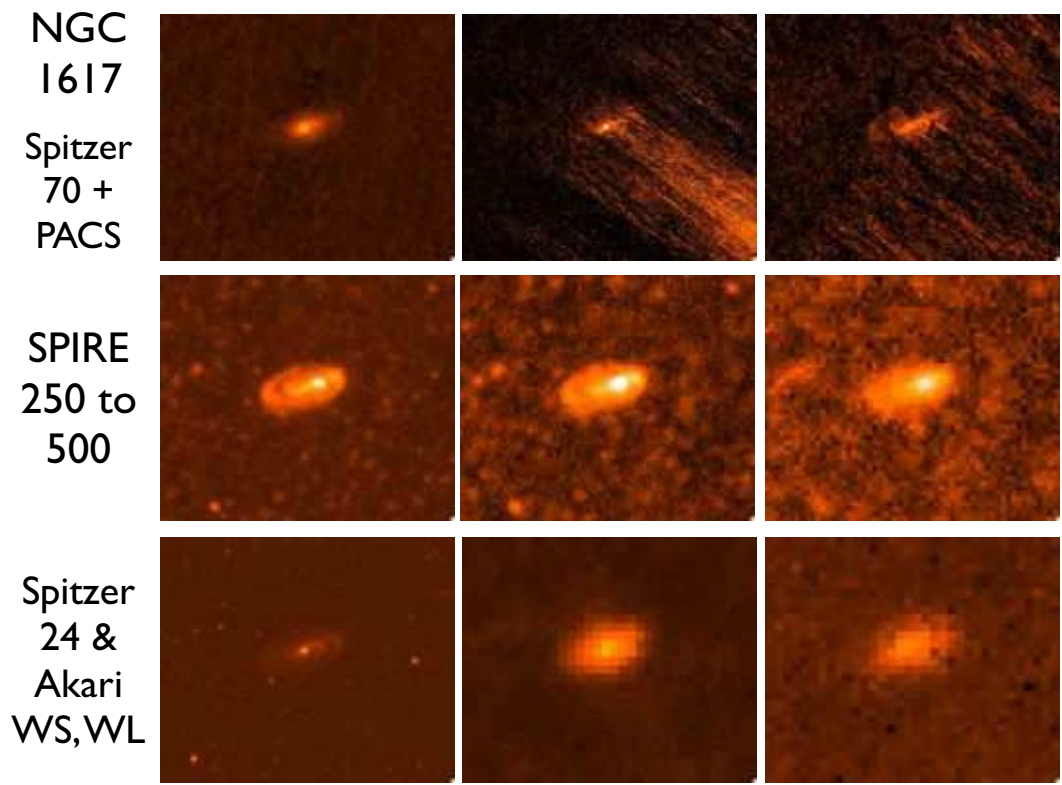}
\caption{The galaxy NGC1617 as revealed from the mid- to far-infrared by ADF-S observations. This source lies at the very edge of the PACS field observed by {\em Herschel}  leading to some of the striping in the PACS images in the top right and top middle, but you can still make out the same structures seen with other instruments .}
\label{fig:ngc1617}
\end{figure*}

An idea of what can be achieved by the combination of this data can be seen in Fig. \ref{fig:ngc1617} which shows a close-up image of the nearby galaxy NGC1617, visible as the bright blob to the bottom right of the field in Fig. \ref{fig:adfs}. One thing that is readily apparent from this comparison is the presence of a blob of bright emission to the right of the galaxy's nucleus in the SPIRE images that is not discernible in the {\em AKARI} or {\em Spitzer} data at shorter wavelengths. This suggests the presence of a region of cold, off nucleus dust that would be missed in observations at wavelengths shorter than $\sim 160 \mu$m. This effect has been seen in other local galaxies and in statistical samples selected by {\em Planck}, and will be discussed in more detail in Rowan-Robinson et al., in prep.

\section{Lensed High Redshift Far-IR Galaxies in ADF-S}

At fluxes lower than $\sim$20 mJy, the number counts for galaxies in the SPIRE bands rise steeply (Clements et al., 2010; Oliver et al., 2010). At brighter fluxes the counts are Euclidean, raising the interesting possibility that a significant number of bright SPIRE sources might be gravitationally lensed. This was first predicted by Negrello et al. (2007), prior to the launch of {\em Herschel} and was rapidly shown to be the case once the first large area survey results arrived (Negrello et al., 2010). Such objects are now the subject of large followup programmes with submm interferometers like the Submillimeter Array (SMA) and ALMA (see eg. Wardlow et al., 2013; Bussmann et al., 2013).

The ADF-S region has been searched for lensing candidates by Bussmann et al. (in prep) using an early reduction of the HerMES SPIRE maps. The search criteria were brightness at 500$\mu$m ($>$ 100mJy in the catalogs available at the time) and an absence of a bright optical or radio counterpart. This selection  is broadly similar to that used by Negrello et al. (2010). Eight lensing candidates in the ADF-S were found which have since been followed up with observations by the Hubble Space Telescope (HST) and ALMA, as part of the larger followup programme. 

\begin{figure}
\centering
\includegraphics[width=85mm]{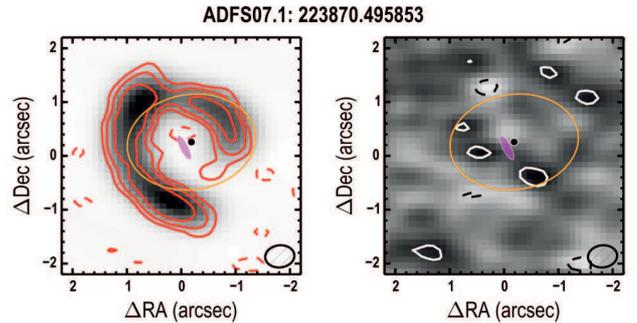}
\caption{Left: ALMA imaging (greyscale and contours) of a strongly lensed far-IR source in the ADF-S. Right: Residuals in the ALMA image after fitting a gravitational lensing model to the image using the {\em uvmcmcfit} code. Details will be presented in Bussmann et al. in prep, code available via github.com/sbussmann/uvmcmcfit. From Bussmann et al., (2015)}
\label{fig:lens}
\end{figure}

The ALMA observations show that all of the eight ADF-S candidates are multiple sources ie. more than one object is responsible for the far-IR emission. In some cases this means that the original sources were in fact blends, not lenses, but there are several possible lens candidates and one very clear lensed source (Bussmann et al., in prep). This is shown in Fig. \ref{fig:lens}, which also shows the residuals after fitting of a source and lens model whose parameters are produced by the {\em uvmcmcfit} code, developed by Bussmann et al. to fit lensing models to interferometric images of lensed far-IR/submm sources. 

\section{Very High Redshift Far-IR Galaxies}

While redshifts have yet to be measured for the ADF-S lensed sources, the redshifts of other lensed {\em Herschel} sources now stretch up to z=5, with one HerMES lensed source with a confirmed redshift of 5.29 (Perez-Fournon et al., in prep) and another, lensed by a cluster of galaxies rather than by a single galaxy, discovered by the {\em Herschel} Lens Survey at z=5.2 (Combes et al., 2012). These sources all have unlensed far-infrared luminosities that would make them at least ULIRGs (ie. L$_{fir} > 10^{12} L_{\odot}$) and possibly HLIRGs (L$_{fir} > 10^{13} L_{\odot}$). The unlensed, or perhaps it would be better to say less lensed given that the optical depth to lensing at redshifts as high as this is $\sim$1, high redshift dusty galaxy population has also been explored by {\em Herschel} through the use of colour as a way of selecting candidate high redshift sources. This is possible because the peak of the far-IR SED of galaxies usually lies at $\sim$100$\mu$m in the emitted frame. A low redshift galaxy would thus appear blue in the SPIRE bands. At redshifts $\sim$ 2 this peak is shifted to the 350$\mu$m band, and at redshifts $>4$ into the 500$\mu$m band. Sources at z$>$4 will therefore have SPIRE SEDs that rise towards the longest wavelength band, and can thus be selected for followup observations. Such programmes are discussed in Dowell et al. (2014). The highest redshift source uncovered by such methods so far, known as HFLS3, is at a redshift of 6.34 (Riechers et al., 2013).

The existence of such sources at such high redshifts is something of a surprise. In fact, essentially none of the current far-IR galaxy population models can reproduce the numbers of such sources seen in the HerMES survey (Dowell et al., 2014). The sources we are seeing at these redshifts have prodigious luminosities, typically $>$10$^{13}$L$_{\odot}$, and so are forming stars very rapidly. While their areal density is low, they account for roughly 10\% of the star formation rate density of the universe that the UV selected populations of galaxies, found in deep optical/near-IR surveys using HST or large ground-based telescopes, are responsible for at these same redshifts. However, if we assume an underlying population of less luminous far-IR galaxies at these same redshifts, of which the sources we are finding are only the high luminosity tip of the iceberg, then simple luminosity function extrapolations mean that such a far-IR population could produce just as many stars as the UV selected sources (Dowell et al., 2014).

The {\em Herschel} data for the ADF-S is also being used for this work, with 17 good candidates selected from the SPIRE colours. These are being followed up using LABOCA (Ivison \& Weiss in prep), and extant data from ASTE/AzTEC (Hatsukade et al., 2011) and AKARI, but no clear redshift determinations are yet available.

\section{ADF-S and Very High Redshift}

The discovery of sources like HFLS3 raises two questions: are there far-IR luminous sources at still higher redshift, and how many lower luminosity sources lie at comparable redshift? These questions are difficult to answer with {\em Herschel} data. Sources at higher redshift than HFLS3 will not be easily detected by {\em Herschel} since the SPIRE bands will be probing ever-shorter wavelengths in the emitted SED. The flux received will thus drop off rapidly with redshift. Lower luminosity sources will also be undetectable by {\em Herschel}. HFLS3 is only about a factor of 2 above the 3$\sigma$ confusion limit. A source with just half the luminosity of HFLS3 lying at the same redshift would thus be essentially invisible to {\em Herschel}.

\begin{figure}
\centering
\includegraphics[width=85mm]{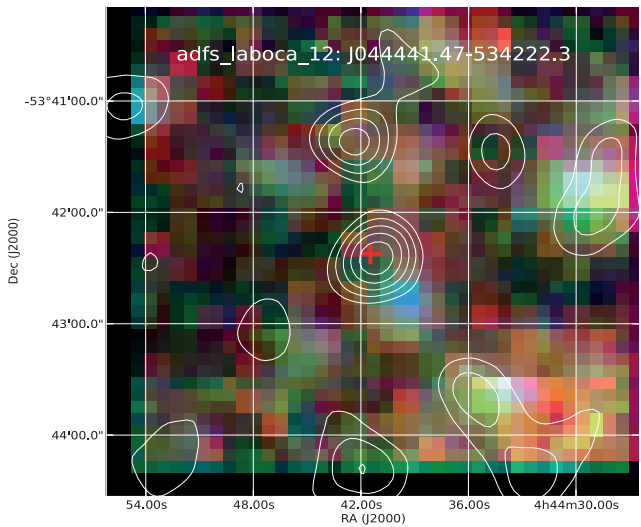}
\caption{Candidate high redshift SPIRE dropout from ADF-S. Colour image is an RGB version of SPIRE with 500, 350 and 250$\mu$ as red, green and blue respectively. Contours are from the 1.1mm ASTE/AzTEC observations of the ADF-S by Hatsukade et al. (2011). As can be seen this bright 1.1mm source lacks a counterpart in the SPIRE bands.}
\label{fig:dropout}
\end{figure}

The solution to this problem has two steps. Firstly, the sources must be detected at longer wavelengths, in the mm or submm bands, using instruments such as SCUBA2 or AzTEC on ASTE. Secondly, they can then be identified as candidate very high redshift sources by their {\em absence} from SPIRE images - they would be, in effect, `SPIRE dropouts' thanks to their high redshift and the shape of the far-IR SED. ADF-S is an ideal testbed for this work since there is extensive {\em Herschel} data and also coverage of 0.25 sq. deg. at 1.1mm (Hatsukade et al., 2011). We have therefore compared the 1.1mm source lists in ADF-S with the SPIRE data and have found a number of candidate SPIRE-dropouts, one of which is shown in Fig. \ref{fig:dropout}. Confirmation of these sources using further submm imaging is underway.

\section{Conclusions}

Many areas of astrophysics benefit from the availability of large fields with imaging at a wide range of wavelengths. The ADF-S is one such field, with data ranging from optical to far-IR and radio. The future for this field is bright, with future surveys due to add breadth and depth to its coverage, and with {\em Herschel} data now available. The potential for new insights into the high redshift universe with the ADF-S is very good.

%


\acknowledgments

I am grateful to the conference organisers for the invitation to give this review and to my many colleagues in the HerMES Consortium for their continuing work on the ADF-S {\em Herschel} observations and many useful discussions.


%


\end{document}